\documentstyle[11pt]{article}
\begin{document}

\title{{\bf Comment on "A holographic model of dark energy and the thermodynamics of a
non-flat accelerated expanding universe", by M.R. Setare and S.
Shafei (JCAP 09 (2006) 011)}}
\author{K. Karami\thanks{E-mail: KKarami@uok.ac.ir}\\
\small{Department of Physics, University of Kurdistan, Pasdaran
St., Sanandaj, Iran}\\}

\maketitle
\begin{abstract}
Authors of ref. \cite{Setare1}, M.R. Setare and S. Shafei (JCAP
{\bf 09} (2006) 011), studied the thermodynamics of a holographic
dark energy model in a non-flat universe enclosed by the apparent
horizon $R_A$ and the event horizon measured from the sphere of
the horizon named $L$. In section 3 in ref. \cite{Setare1},
Authors showed that for $R_A$ the generalized second law of
thermodynamics is respected, while for $L$ it is satisfied for the
special range of the deceleration parameter. Here we present that
their calculations for $R_A$ should be revised. Also we show that
their conclusion for $L$ is not true and the generalized second
law is hold for the present time independently of the deceleration
parameter. Also if we take into account the contribution of dark
matter in the generalized second law which is absent in ref.
\cite{Setare1}, then the generalized second law for $L$ is
violated for the present time.
\end{abstract}

\clearpage

\section{Second law of thermodynamics}
The apparent horizon, $R_A$, in a non-flat universe is given by
\cite{Cai}
\begin{equation}
R_A=H^{-1}(1+\Omega_{k})^{-1/2},\label{ah}
\end{equation}
whereas Authors of ref. \cite{Setare1} considered $R_A=H^{-1}$ for a
non-flat universe. For the flat case, i.e. $\Omega_{k} = 0$, the
apparent horizon is same as the Hubble horizon. Therefore Eqs. (30)
and (31) in ref. \cite{Setare1} must be corrected, respectively, as
follows
\begin{equation}
{\rm d}S=\pi
(1+3\omega_{\Lambda}\Omega_{\Lambda}H^2R_{A}^2)R_{A}{\rm d}R_{A},
\end{equation}
\begin{eqnarray}
\frac{{\rm d}S}{{\rm
d}x}&=&-\pi(1+\Omega_{k}+3\omega_{\Lambda}\Omega_{\Lambda})H^{-2}\frac{\Big(H^{-1}\frac{{\rm
d}H}{{\rm
d}x}-\Omega_{k}\Big)}{(1+\Omega_{k})^3},\nonumber\\&=&-\pi
\Big(1+\Omega_{k}-\Omega_{\Lambda}-\frac{2}{c}\Omega_{\Lambda}^{3/2}\cos{y}\Big)H^{-2}\frac{\Big(H^{-1}\frac{{\rm
d}H}{{\rm
d}x}-\Omega_{k}\Big)}{(1+\Omega_{k})^3},\nonumber\\&=&-2\pi
qH^{-2}\frac{\Big(H^{-1}\frac{{\rm d}H}{{\rm
d}x}-\Omega_{k}\Big)}{(1+\Omega_{k})^3},
\end{eqnarray}
where $q$ is the deceleration parameter and from Eq. (32) in ref.
\cite{Setare1} it can be rewritten without approximation as
\begin{equation}
q=-1-H^{-1}\frac{{\rm d}H}{{\rm
d}x}=-\frac{\Omega_{\Lambda}^{3/2}\cos
y}{c}+\frac{1-\Omega_{\Lambda}}{2(1-a\gamma)}
=\frac{1}{2}\Big(1+\Omega_{k}-\Omega_{\Lambda}-\frac{2}{c}\Omega_{\Lambda}^{3/2}\cos{y}\Big).\label{qexact1}
\end{equation}
Using Eq. (\ref{ah}) one can obtain
\begin{equation}
R_A{\rm d}R_A=-H^{-2}\frac{\Big(H^{-1}\frac{{\rm d}H}{{\rm
d}x}-\Omega_{k}\Big)}{(1+\Omega_{k})^2}{\rm d}x,
\end{equation}
whereas in ref. \cite{Setare1}, $R_A{\rm d}R_A=-H^{-3}({\rm d}H/{\rm
d}x){\rm d}x$.

Using Eq. (\ref{ah}), the corrections of Eqs. (33) and (34) in ref.
\cite{Setare1} are obtained, respectively, as
\begin{equation}
\frac{{\rm d}S_{A}}{{\rm d}x}=-2\pi
H^{-2}\frac{\Big(H^{-1}\frac{{\rm d}H}{{\rm
d}x}-\Omega_{k}\Big)}{(1+\Omega_{k})^2},
\end{equation}
\begin{equation}
\frac{\rm d}{{\rm d}x}(S+S_{A})=\frac{2\pi
H^{-2}}{(1+\Omega_{k})^3}\Big(H^{-1}\frac{{\rm d}H}{{\rm
d}x}-\Omega_{k}\Big)^2.
\end{equation}
For the event horizon measured from the sphere of the horizon
named $L$, from Eq. (29) in ref. \cite{Setare1} and using
$\rho_{\Lambda}=\frac{3c^2}{8\pi}L^{-2}$ and $E=\frac{4}{3}\pi
L^3\rho=\frac{1}{2}c^2L$, Eq. (40) in ref. \cite{Setare1} is
corrected as
\begin{equation}
\frac{{\rm d}S}{{\rm d}x}=\frac{-\pi c^4
}{H^2\Omega_{\Lambda}^2}\Big(\frac{\Omega_{\Lambda}+3\omega_{\Lambda}\Omega_{\Lambda}}{H}\frac{{\rm
d}H}{{\rm
d}x}+\frac{1+3\omega_{\Lambda}}{2}\Omega_{\Lambda}^{'}\Big),\label{dSdx1}
\end{equation}
where $\Omega_{\Lambda}^{'}$ is given by Eq. (22) in ref.
\cite{Setare1} and can be rewritten as
\begin{equation}
\Omega_{\Lambda}^{'}=\frac{2}{c}\Omega_{\Lambda}^{3/2}\cos
y+2q\Omega_{\Lambda}.\label{OmegaLp}
\end{equation}
Due to have a correct dimension, the holographic dark energy (DE)
density, $\rho_{\Lambda}$, given by Eq. (5) in ref. \cite{Setare1}
should be corrected as $\rho_{\Lambda}=\frac{3c^2}{8\pi}L^{-2}$,
where we take $G=1$.

Using Eq. (\ref{dSdx1}) for the evolution of entropy of the DE
inside the universe enclosed by the horizon $L$ and Eq. (41) in
ref. \cite{Setare1} for the evolution of the geometric entropy of
the horizon, one can obtain
\begin{equation}
\frac{\rm d}{{\rm d}x}(S+S_{\rm L})=\frac{\pi c^4
}{H^2\Omega_{\Lambda}^2}\Big(\Omega_{\Lambda}+3\omega_{\Lambda}\Omega_{\Lambda}+\frac{2\Omega_{\Lambda}}{c^2}\Big)
\Big(1+q-\frac{\Omega_{\Lambda}^{'}}{2\Omega_{\Lambda}}\Big).\label{dStot1}
\end{equation}

Here we would like to correct Eq. (42) in ref. \cite{Setare1}
using the approximation given by Eq. (32) in ref. \cite{Setare1}
for the deceleration parameter $q$. This procedure is same as that
used by Authors of ref. \cite{Setare1}. To do this, from Eq. (32)
in ref. \cite{Setare1} we have
\begin{equation}
\frac{\Omega_{\Lambda}^{3/2}\cos
y}{c}\simeq-q+\frac{1-\Omega_{\Lambda}}{2},\label{qapprox}
\end{equation}
then substituting Eq. (\ref{qapprox}) in both Eq. (9) in ref.
\cite{Setare1} and Eq. (\ref{OmegaLp}) we obtain
\begin{equation}
3\omega_{\Lambda}\Omega_{\Lambda}\simeq 2q-1,\label{approx1}
\end{equation}
\begin{equation}
\Omega_{\Lambda}^{'}\simeq(2q-1)(\Omega_{\Lambda}-1).\label{approx2}
\end{equation}
Substituting Eqs. (\ref{approx1}) and (\ref{approx2}) in Eq.
(\ref{dStot1}), we get the corrected form of Eq. (42) in ref.
\cite{Setare1} with using the approximation given by Eq. (32) in
ref. \cite{Setare1} as
\begin{equation}
\frac{\rm d}{{\rm d}x}(S+S_{\rm L})=\frac{\pi c^4
}{H^2\Omega_{\Lambda}^2}\left\{\Big(2q+\frac{2\Omega_{\Lambda}}{c^2}+\Omega_{\Lambda}-1\Big)
\Big[\Big(\frac{1-\Omega_{\Lambda}}{2\Omega_{\Lambda}}\Big)(2q-1)+1+q\Big]\right\},\label{GSL11}
\end{equation}
or
\begin{eqnarray}
\frac{\rm d}{{\rm d}x}(S+S_{\rm L})=\frac{\pi c^4
}{H^2\Omega_{\Lambda}^2}\left\{(1+q)\Big(2q+\frac{2\Omega_{\Lambda}}{c^2}\Big)
+\Big(\frac{1-\Omega_{\Lambda}}{\Omega_{\Lambda}}\Big)(2q-1)\Big(1+q+\frac{\Omega_{\Lambda}}{c^2}\Big)
\right.\nonumber\\\left.-(1-\Omega_{\Lambda})
\Big[(2q-1)\Big(\frac{3-\Omega_{\Lambda}}{2\Omega_{\Lambda}}\Big)+1+q\Big]\right\}.\label{GSL12}
\end{eqnarray}
Taking $\Omega_{\Lambda}=0.73$ and $c=1$ given by ref.
\cite{Setare1} for the present time, Eq. (\ref{GSL11}) gives
\begin{eqnarray}
\frac{\rm d}{{\rm d}x}(S+S_{\rm L})&=&\frac{5.89528
}{H^2}(0.815068+1.36986q)(1.19+2q),\nonumber\\&=&\frac{16.1515
}{H^2}(q+0.5949)^2\geq 0,\label{GSL13}
\end{eqnarray}
which compared to Eq. (43) in ref. \cite{Setare1} shows that in
contrary to the conclusion of Authors of ref. \cite{Setare1}, the
generalized second law (GSL) of thermodynamics for the holographic
DE in a non-flat universe enveloped by the horizon $L$ is
satisfied for the present time independently of the deceleration
parameter $q$.

Here we would like to correct again Eq. (42) in ref. \cite{Setare1}
but this time without using the approximation given by Eq. (32) in
ref. \cite{Setare1} for the deceleration parameter $q$. To do this,
from the exact relation for $q$ given by Eq. (32) in ref.
\cite{Setare1} we have
\begin{equation}
\frac{\Omega_{\Lambda}^{3/2}\cos
y}{c}=-q+\frac{1-\Omega_{\Lambda}}{2(1-a\gamma)},\label{qexact2}
\end{equation}
then substituting Eq. (\ref{qexact2}) in both Eq. (9) in ref.
\cite{Setare1} and Eq. (\ref{OmegaLp}) we obtain
\begin{equation}
3\omega_{\Lambda}\Omega_{\Lambda}=2q-\Omega_{\Lambda}-\Big(\frac{1-\Omega_{\Lambda}}{1-a\gamma}\Big),\label{exact1}
\end{equation}
\begin{equation}
\Omega_{\Lambda}^{'}=(\Omega_{\Lambda}-1)\Big(2q-\frac{1}{1-a\gamma}\Big).\label{exact2}
\end{equation}
Substituting Eqs. (\ref{exact1}) and (\ref{exact2}) in Eq.
(\ref{dStot1}), we get the corrected form of Eq. (42) in ref.
\cite{Setare1} without using the approximation given by Eq. (32) in
ref. \cite{Setare1} as
\begin{equation}
\frac{\rm d}{{\rm d}x}(S+S_{\rm L})=\frac{\pi c^4
}{H^2\Omega_{\Lambda}^2}\left\{\Big[2q+\frac{2\Omega_{\Lambda}}{c^2}-\Big(\frac{1-\Omega_{\Lambda}}{1-a\gamma}\Big)\Big]
\Big[\Big(\frac{1-\Omega_{\Lambda}}{2\Omega_{\Lambda}}\Big)\Big(2q-\frac{1}{1-a\gamma}\Big)+1+q\Big]\right\},\label{GSL21}
\end{equation}
or
\begin{eqnarray}
\frac{\rm d}{{\rm d}x}(S+S_{\rm L})=\frac{\pi c^4
}{H^2\Omega_{\Lambda}^2}\left\{(1+q)\Big(2q+\frac{2\Omega_{\Lambda}}{c^2}\Big)
+\Big(\frac{1-\Omega_{\Lambda}}{\Omega_{\Lambda}}\Big)\Big(2q-\frac{1}{1-a\gamma}\Big)\Big(1+q+\frac{\Omega_{\Lambda}}{c^2}\Big)
\right.\nonumber\\\left.-(1+q)\Big(\frac{1-\Omega_{\Lambda}}{1-a\gamma}\Big)
-\Big(2q-\frac{1}{1-a\gamma}\Big)\Big(\frac{1-\Omega_{\Lambda}}{2\Omega_{\Lambda}}\Big)
\Big(2+\frac{1-\Omega_{\Lambda}}{1-a\gamma}\Big)\right\}.\label{GSL22}
\end{eqnarray}
Note that Eq. (\ref{GSL21}) reduces to Eq. (\ref{GSL11}) when
$\gamma:=\Omega_{k}^{0}/\Omega_{\rm m}^{0}$ goes to zero.

For $\Omega_{\Lambda}=0.73$, $c=1$ and $\gamma\sim0.04$ given by
ref. \cite{Setare1} for the present time $a=1$, Eq. (\ref{GSL21})
gives
\begin{eqnarray}
\frac{\rm d}{{\rm d}x}(S+S_{\rm L})&=&\frac{5.89528
}{H^2}(0.807363+1.36986q)(1.17875+2q),\nonumber\\&=&\frac{16.1515
}{H^2}(q+0.5894)^2\geq 0,\label{GSL23}
\end{eqnarray}
which shows that same as the result obtained by Eq. (\ref{GSL13})
and in contrary to the conclusion of Authors of ref.
\cite{Setare1}, the GSL for the horizon $L$ is satisfied again for
the present time independently of the deceleration parameter $q$.

Although Authors of ref. \cite{Setare1} have considered both of the
DE and dark matter (DM) in their model (see Eq. (18) in ref.
\cite{Setare1}), they have not taken into account the contribution
of the DM in the GSL (see again Eq. (42) in ref. \cite{Setare1}).
Therefore to complete the calculations, the contribution of the
entropy of the DM should be considered in the GSL. To do this, from
Eq. (29) in ref. \cite{Setare1} and using $P_{\rm m}=0$ and $E_{\rm
m}=\frac{4}{3}\pi L^3\rho_{\rm m}$, the evolution of entropy of the
DM inside the universe enclosed by the horizon $L$ is obtained as
\begin{equation}
\frac{{\rm d}S_{\rm m}}{{\rm d}x}=\frac{3\pi c^4
}{H^2\Omega_{\Lambda}^2}\Big(q-\frac{\Omega_{\Lambda}^{'}}{2\Omega_{\Lambda}}\Big)\Omega_{\rm
m}.\label{dSm}
\end{equation}
Finally, using Eqs. (\ref{dStot1}) and (\ref{dSm}), the GSL due to
different contributions of the DE, DM and horizon $L$ can be
obtained as
\begin{equation}
\frac{\rm d}{{\rm d}x}(S+S_{\rm L}+S_{\rm m})=\frac{\pi c^4
}{H^2\Omega_{\Lambda}^2}\left\{\Big(\Omega_{\Lambda}+3\omega_{\Lambda}\Omega_{\Lambda}+\frac{2\Omega_{\Lambda}}{c^2}+3\Omega_{\rm
m}\Big)
\Big(1+q-\frac{\Omega_{\Lambda}^{'}}{2\Omega_{\Lambda}}\Big)-3\Omega_{\rm
m}\right\}.\label{dStot2}
\end{equation}
Using Eq. (9) in ref. \cite{Setare1} and Eq. (\ref{OmegaLp}), we can
rewrite Eq. (\ref{dStot2}) as
\begin{eqnarray}
\frac{{\rm d}}{{\rm d}x}(S+S_{\rm L}+S_{\rm m})=\frac{\pi
c^4}{H^2\Omega_{\Lambda}^2}\left\{\frac{2\Omega_{\Lambda}}{c^2}\Big(1-\frac{\sqrt{\Omega_{\Lambda}}}{c}\cos{y}\Big)\Big(1-c\sqrt{\Omega_{\Lambda}}\cos{y}\Big)
\right.\nonumber\\\left.-3(1+\Omega_{k}-\Omega_{\Lambda})\frac{\sqrt{\Omega_{\Lambda}}}{c}\cos{y}\right\},\label{dStot3}
\end{eqnarray}
which is same as Eq. (1.6) in ref. \cite{Karami} with $b^2=0$.

Taking $\Omega_{\Lambda}=0.73$, $\Omega_{k}=0.01$, $c=1$
\cite{Setare1} and $\cos y=0.99$ \cite{Karami} for the present time,
Eq. (\ref{dStot3}) gives
\begin{equation}
\frac{\rm d}{{\rm d}x}(S+S_{\rm L}+S_{\rm
m})=-\frac{3.98421}{H^2}<0,
\end{equation}
which shows that same as the result obtained by \cite{Karami} and in
contrary to the conclusions of Eqs. (\ref{GSL13}) and (\ref{GSL23}),
the GSL is violated at the present time for a non-flat universe
containing the holographic DE and DM and enveloped by the event
horizon measured from the sphere of the horizon named $L$.

\end{document}